\documentclass[sigconf]{acmart}
\usepackage{framed,xcolor,graphicx}
\definecolor{shadecolor}{cmyk}{0,0,0,.1}
\usepackage{enumitem}

\AtBeginDocument{%
  }

\copyrightyear{2025}
\acmYear{2025}
\setcopyright{rightsretained}
\acmConference[CHI EA '25]{Extended Abstracts of the CHI Conference on Human Factors in Computing Systems}{April 26-May 1, 2025}{Yokohama, Japan}
\acmBooktitle{Extended Abstracts of the CHI Conference on Human Factors in Computing Systems (CHI EA '25), April 26-May 1, 2025, Yokohama, Japan}\acmDOI{10.1145/3706599.3719938}
\acmISBN{979-8-4007-1395-8/2025/04}

\begin{document}

\title{From Everyday Technologies to Augmented Reality: An Autoethnographic Study of Presence and Engagement}

\author{Tram Thi Minh Tran}
\email{tram.tran@sydney.edu.au}
\orcid{0000-0002-4958-2465}
\affiliation{Design Lab, Sydney School of Architecture, Design and Planning,
  \institution{The University of Sydney}
  \city{Sydney}
  \state{NSW}
  \country{Australia}
}

\renewcommand{\shortauthors}{Tran}



\begin{abstract} 

Digital technologies are reshaping how people experience their surroundings, often pulling focus toward virtual spaces and making it harder to stay present and engaged. Wearable augmented reality (AR), by embedding digital information into the physical world, may further immerse users in digital layers. Yet paradoxically, it also holds the potential to support presence and engagement. To explore this possibility, this study adopts an autoethnographic approach, providing a first-person perspective on how everyday technologies shape real-world engagement. Over four weeks, 20 experiences were documented, capturing interactions with phones, laptops, and fitness trackers in various contexts. The findings reveal nuanced patterns of technology use and propose design implications for wearable AR, emphasising its potential for personalised, context-aware interventions that support meaningful real-world connection. This work contributes to the discourse on digital well-being, suggesting that wearable AR can evolve beyond digital augmentation to help users reconnect with their surroundings.

\end{abstract}



\begin{teaserfigure}
  \includegraphics[width=\textwidth]{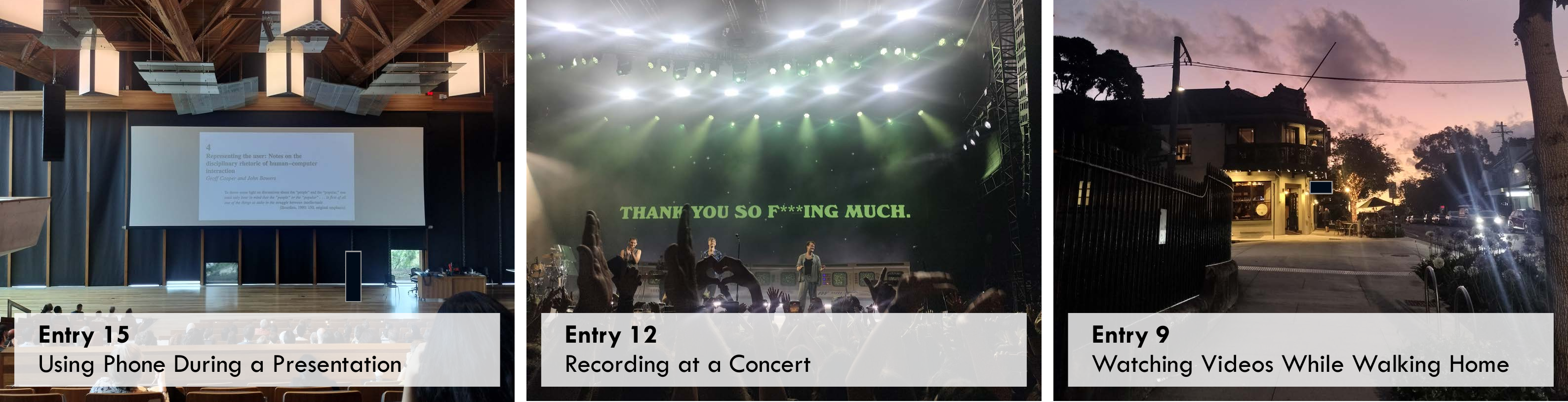}
  \caption{Snapshots from selected entries in the autoethnographic study, depicting everyday contexts where technology influenced presence and engagement. These contexts form the foundation for exploring design implications for wearable augmented reality.}
  \Description{A collage of three images, each capturing a moment from the author’s perspective where technology influenced their presence and engagement. The first image (Entry 15) shows a lecture hall during a presentation, with the author using their phone while a slide is projected at the front. The second image (Entry 12) captures a concert scene, where the author records the performance. The third image (Entry 9) depicts a quiet street at dusk, taken while the author watches videos on their phone while walking home. These snapshots illustrate everyday interactions with technology, forming the foundation for exploring design implications for wearable augmented reality.}
  \label{fig:teaser}
\end{teaserfigure}

\maketitle

\section{Introduction}

In today’s hyper-connected world, staying present is increasingly challenging. Notifications buzz, devices demand attention, and our focus is frequently divided between the digital and physical~\cite{landon2024stolen}. As an interaction designer fascinated by emerging technologies, I find myself questioning whether these tools truly serve us. While their potential to connect and inform is undeniable, they often leave me feeling disconnected from the world around. This tension has led to a growing interest in how technologies can mediate, rather than fragment presence.

Wearable augmented reality (AR)~\cite{azuma2019road, billinghurst2021grand}, a technology that overlays digital information onto the physical world, presents a compelling paradox: could it enhance presence and engagement in the real world rather than amplifying digital immersion? This study takes inspiration from Keiichi Matsuda’s Hyper-Reality concept film\cite{matsuda2016hyperreality}, which critiques the overwhelming saturation of digital layers in everyday life. It explores whether wearable AR could take a different trajectory—one that enhances real-world connection instead of amplifying detachment.

To investigate this potential, I conducted an autoethnographic study~\cite{lucero2018living, kaltenhauser2024playing} to examine how everyday technologies—including phones, laptops, and fitness trackers—mediate presence and engagement with the surrounding environment. Over four weeks, 20 entries captured specific interactions, emotional responses, and reflections across diverse contexts, ranging from professional settings to personal activities. By focusing on lived experiences, this approach allows for a nuanced exploration of how current technologies influence presence and what lessons they offer for future wearable AR design.

This research is guided by two key questions:
\begin{itemize}[topsep=0pt]
\item \textbf{RQ1}: How do everyday technologies influence my presence and engagement with my surroundings in various contexts?
\item \textbf{RQ2}: How can wearable AR be designed to support presence and engagement?
\end{itemize}

This study contributes: (1) an analysis of how everyday technologies mediate presence and engagement with the physical environment, and (2) design implications for wearable AR systems that emphasise context-aware, real-world engagement. By critically reflecting on everyday interactions, this research expands the discourse on the role of technology in fostering digital well-being. It envisions a future where wearable AR evolves beyond a tool for digital augmentation to become a bridge that reconnects users with their physical surroundings.

\section{Related Work}

\subsection{Digital Wellbeing and Design for Disengagement}

Digital wellbeing and design for disengagement have emerged as responses to growing concerns about technology dependency and its impact on mental health, attention, and quality of life~\cite{obrien2022rethinking, cecchinato2019designing}. Historically, consumer technologies were designed to maximise user engagement, often at the expense of agency and wellbeing. In recent years, however, technology companies have introduced features under the umbrella of digital wellbeing~\cite{android2024wellbeing}, aimed at addressing overuse by incorporating friction into digital interactions, encouraging intentional breaks, and raising awareness of usage patterns. Examples include tools for monitoring and managing screen time~\cite{tiktok_screen_time, google_screen_time}, focus modes to minimise distractions~\cite{apple_focus_mode, android2024wellbeing}, gamified approaches to limit device use~\cite{forest_app}, and bedtime routines to support transitions to rest~\cite{android2024wellbeing}.

Design for disengagement aims to empower users to regain control over their technology habits, offering alternatives to extreme measures like digital detoxes or complete abstinence from technology~\cite{radtke2022digital, syvertsen2020}. However, these strategies are predominantly reactive, focusing on mitigating the negative effects of digital engagement rather than reimagining how technology might actively foster presence and engagement with the physical world.

\subsection{Augmented Reality and Its Potential for Digital Wellbeing}

Wearable AR has been applied in both specialised domains—such as industrial training and medical procedures—and everyday contexts, including entertainment and productivity~\cite{tran2023wearable}. With its compact form factor and envisioned ubiquity as smartglasses, wearable AR is designed to seamlessly integrate into daily life, offering hands-free, context-aware interactions that distinguish it from traditional screen-based technologies~\cite{azuma2019road, billinghurst2021grand}. While much research on AR has focused on enhancing user experiences by overlaying relevant digital information onto the physical world, its potential as a tool for supporting digital wellbeing remains underexplored. Unlike smartphones, which often pull users away from their surroundings, wearable AR has the potential to facilitate a more fluid integration of digital and real-world interactions, adapting to the user’s situational context.

However, the ability of AR to embed digital content directly into the user’s field of view also raises concerns about amplifying the same isolating effects seen with existing technologies. Persistent overlays or excessive digital distractions could further fragment attention, making users less engaged with their surroundings or the people around them~\cite{matsuda2016hyperreality}. These potential drawbacks highlight the need for thoughtful AR design that prioritises meaningful real-world engagement while minimising the risk of overuse or detachment.

This research positions wearable AR not as a tool to be disengaged from, but as a medium that can enhance presence and engagement in everyday life. Rather than reinforcing digital immersion, well-designed AR experiences could redirect attention toward real-world details, helping users remain situated in their physical environment. 

\section{Method}

This study adopts an \textit{autoethnographic} approach, a qualitative research method that leverages the researcher's first-hand experiences and reflexive analysis~\cite{adams2016handbook, desjardins2021introduction, hook2019sample}. Autoethnography is increasingly utilised in Human-Computer Interaction to provide rich, contextual, and nuanced insights into how technologies intersect with everyday practices, emotions, and behaviours~\cite{kaltenhauser2024playing, lucero2018living, o2014gaining}.

\subsection{Positionality and Ethical Considerations}
As both a researcher and participant in this autoethnographic study, I draw on my experiences as a design researcher with a professional interest in emerging technologies and a personal awareness of the challenges posed by digital distractions. My background in interaction design informs my focus on how technologies—particularly wearable AR—might support real-world presence and engagement rather than reinforcing detachment. While my reflections are shaped by the devices and tools I interact with daily, this study seeks to situate these experiences within broader discussions of digital wellbeing and AR design. This research is limited solely to my personal experiences and does not involve other participants or identifiable individuals, ensuring that ethical considerations regarding privacy and consent are met.

\subsection{Data Collection Through Self-Observation}

To explore how wearable AR might be designed to support presence and engagement, I systematically documented my interactions with digital technologies in everyday contexts. This self-observation provided insights into how current technologies shape presence and engagement, informing design opportunities for wearable AR that prioritise real-world connection.

\subsubsection{Scope of Technology Interactions}

I focused on devices that frequently mediate my perception of and connection to my environment. These included laptops and desktops, which I primarily used for reading, work, and browsing; mobile phones, utilised for applications such as navigation, photography, social media, and music streaming; and fitness trackers, used for step counting and activity monitoring. These interactions were recorded across diverse everyday environments, including outdoor walks, commuting, social gatherings, home-based activities, and events such as conferences or concerts (see \autoref{fig:teaser}). This diversity ensured a broad exploration of how technology mediates presence and engagement in daily life.

\subsubsection{Reflective Journals}

Data collection spanned four weeks (13 November – 16 December 2024), covering 13 days and resulting in 20 documented experiences. Rather than requiring a journal entry each day, I recorded experiences when notable interactions with technology occurred, aligning with the reflective and emergent nature of autoethnographic inquiry. Each experience was systematically documented with the following details:

\begin{itemize} \item \textit{Technology}: The device used and its intended purpose.
\item \textit{Context}: The physical setting, time of day, and situational details.
\item \textit{Emotional response}: Emotional states or reactions during the experience.
\item \textit{Impact on presence}: Whether the technology enhanced presence and engagement or caused distraction.
\end{itemize}

An example entry is provided in \autoref{tab:entry1_laptop}. The complete dataset is available in \autoref{recorded-xp}. At the end of each day, I expanded my entries with more in-depth reflections, guided by the following prompts:  

\begin{itemize}
    \item Did any technology foster a sense of connection to my environment? How?
    \item Were there moments when technology hindered my engagement with my surroundings?
    \item How did my emotional responses change across different technologies throughout the day?
    \item Are there aspects of this experience that could inform wearable AR design to better support presence and engagement in the physical world?
\end{itemize}

\begin{table*}[!ht]
\centering
\small
\caption{Example Entry - Laptop Use Before Meeting}
\label{tab:entry1_laptop}
\begin{tabular}{p{3cm}p{11cm}}
\toprule
\textbf{Date:} & \textbf{Wednesday, 13 November 2024} \\ 
\midrule
\textbf{Device/Application} & Laptop \\ 
\textbf{Purpose} & Browsing through various applications (Slack, Outlook, Overleaf) without a specific focus. \\ 
\textbf{Context} & Waiting for a meeting to start as one participant is delayed due to traffic. \\ 
\textbf{Emotional Response} & Bored; seeking something to engage with, but not focused on anything specific. \\ 
\textbf{Impact on Presence} & I wasn’t fully engaged with the conversations happening around me. At times, I set my laptop aside to look at the people speaking, but my attention didn’t shift towards them meaningfully. \\ 
\bottomrule
\end{tabular}
\end{table*}

\subsection{Data Analysis}

I systematically examined 20 journal entries using reflexive thematic analysis~\cite{braun2006thematic}. The analysis followed an iterative, inductive approach. First, I familiarised myself with the data by reading through it multiple times, identifying recurring patterns in presence, engagement, and technology use. Next, I performed initial open coding, labelling key aspects of each experience. These codes were then grouped into broader themes, capturing overarching trends across experiences. These themes are presented in two parts within the results section: \textit{Technology in Everyday Engagements}, which illustrates the diverse ways digital technologies mediate my presence and engagement, and \textit{Design Implications for Wearable AR}, which synthesises insights from these reflections to propose considerations for AR design.

\section{Results}

\subsection{Technology in Everyday Engagements}

\textcircled{\small{1}} \textbf{Mindless scrolling and boredom}: Technology often acted as a distraction or coping mechanism, preventing meaningful connections to the environment. For instance, browsing a laptop during a meeting delay filled time without purpose, while morning phone scrolling flooded me with information, leaving no mental space to acknowledge the start of a new day—how the room felt, the weather outside, or the atmosphere around me. At night, aimless browsing became a way to cope with sleeplessness but instead kept me absorbed in endless content, making it harder to relax or fall asleep.

\textcircled{\small{2}} \textbf{Intentional retreat from reality}: At times, technology provided a deliberate escape from discomfort. During a pitching event dinner, my phone served as a social shield, helping me appear busy and avoid awkwardness. Similarly, earphones and music blocked out disruptive construction noise near the office, helping me withdraw entirely from the noisy surroundings. While these interactions offered a sense of control, they came at the cost of engagement with the physical space.

\textcircled{\small{3}} \textbf{Trade-off between documentation and presence}: Technology enabled documentation but often fragmented real-time presence. Recording clips at a concert preserved moments for later but limited physical participation, such as dancing. Similarly, taking photos of slides and Googling information during a conference divided attention, reducing immersion in the presentation.

\textcircled{\small{4}} \textbf{Frustrations with technology being too unobtrusive or absent}: The Fitbit, typically a quiet background tool, became frustrating when its battery ran out at inconvenient moments. Although it sent notifications, they often arrived when I was unable to charge it, disrupting my activity-tracking routine. At a classical concert, the absence of technology initially allowed for deep immersion in the performance, enhancing focus on the music and the acoustics of the hall. However, during the encore, the lack of listed pieces caused a moment of mental distraction as I tried to memorise melodies and guess the composer for later searches. 

\textcircled{\small{5}} \textbf{Technology complementing real-world activities}: In some instances, technology enhanced real-world experiences without causing detachment. Listening to music while walking created a `main character vibe,' enriching the walk emotionally without disconnecting from the surroundings entirely. Likewise, reading a novel on my phone while having breakfast provided a harmonious and joyful moment, where the two activities complemented each other seamlessly.

\textcircled{\small{6}} \textbf{Mediating connections to the environment}: Technology sometimes facilitated engagement with the physical world. While studying Japanese using ChatGPT's speech mode, the AI prompted me with questions like \textit{`What do you see during your walk?'}, encouraging me to describe my surroundings in Japanese. In a different way, playing music on a speaker while reading shaped the atmosphere, making the environment feel nostalgic and immersive.

\textcircled{\small{7}} \textbf{Reality pulling me back}: Unexpected physical-world cues often interrupted digital immersion, offering moments of reconnection. On the way home, the sound of crickets and the glow of the sunset led me to pause scrolling, remove my earphones, and become more aware of the breeze, fading light, and ambient sounds. A similar occurrence took place at a laundromat. While reading and listening to music, a cool summer breeze from the open door momentarily redirected my focus to the passing traffic and changing atmosphere of the season.

\subsection{Design Implications for Wearable AR} 

Drawing from the diverse ways digital technologies mediate my presence and engagement (\textcircled{\small{1}}–\textcircled{\small{7}}), this part of the analysis outlines key design considerations for wearable AR, envisioning how it could be designed to better support real-world connection.


\subsubsection{Balancing Technology Immersion}

With its contextual awareness and ability to recognise user habitual behaviours, AR could detect aimless scrolling or passive interaction~\textcircled{\small{1}} and offer gentle nudges to mindful break, such as observing their environment or taking a short reflective pause. For example, during habitual nighttime browsing, AR might suggest restful alternatives like guided breathing or audio reflections, tailored to the user's mental state. Additionally, because AR operates within a user’s immediate environment, it can provide subtle disconnection cues, such as dimming colours or adding ambient visuals (e.g., a sunset fade), to encourage natural transitions from overstimulation to rest.

In contexts like a networking event, where users might seek to withdraw to avoid social discomfort~\textcircled{\small{2}}, AR could provide subtle, low-friction tools to balance escapism with situational awareness. For instance, ambient overlays could display attendee profiles, conversation topics, or shared interests, enabling users to navigate the social environment more confidently. These features would support manageable interactions, allowing users to stay engaged without feeling overwhelmed or completely withdrawing. In noisy work environments, AR could enhance noise-cancellation systems by incorporating awareness prompts, ensuring disruptive sounds are filtered out while critical cues, such as doorbells or approaching footsteps, remain detectable.

\begin{shaded}
\noindent AR can help users navigate the dual challenge of disengaging from technology when it becomes mindless and supporting its use when intentional and purposeful. 
\end{shaded}

\subsubsection{Balancing Digital Tasks and Physical Presence}


By leveraging simple gesture-based interactions, AR could enable hands-free content capture without requiring extensive device handling, such as for recording slides during presentations. Additionally, when unfamiliar terms or concepts arise, AR could provide real-time contextual lookups, allowing users to access relevant information seamlessly. Users could trigger lookups through subtle interactions, such as a quick gaze hold on a word, a double blink, or a small hand gesture. Captured content could be automatically organised, summarised, and resurfaced in relevant contexts, ensuring that information remains easily accessible rather than becoming forgotten or overwhelming. This auto-documentation approach could help users stay physically engaged, minimising disruptions caused by holding devices or switching tasks~\textcircled{\small{3}}.

To avoid frustration with too unobstrusive technology~\textcircled{\small{4}}, context-aware notifications and task reminders, could be further refined by factoring in user location and context. For instance, a battery notification would be relevant only when the charger is nearby, avoiding unnecessary interruptions when at home or away from the office. For seamless information retrieval, AR could incorporate a passive recognition feature that automatically detects and discreetly displays minimal details—such as composer and piece titles at a concert—without requiring explicit user input.


\begin{shaded}
\noindent AR has the potential to make interactions with the real world more seamless and efficient, ensuring users can balance digital tasks with physical presence. 
\end{shaded}

\subsubsection{Drawing Attention to the Real World}

AR can play a complementary role~\textcircled{\small{5}} by providing ambient, non-intrusive enhancements to activities. For example, AR could provide subtle safety prompts to stay attentive in specific contexts, such as intersections while walking with music. AR could adjust lighting and colour tones to suit a user’s environment during reading, creating a more harmonious and pleasant experience.

AR can mediate connections to the physical environment~\textcircled{\small{6}} by enhancing the right moments or remaining in the background when appropriate. For example, during activities such as walking or practising a language, AR could overlay real-time contextual information, like translations or spelling corrections, directly onto the environment using conformal overlays. This approach supports immersive learning without disrupting the flow of engagement. In certain situations, AR could blend into the background by adapting music or soundscapes to complement the physical environment. Since music often plays a supporting role in activities like work or relaxation, AR could automatically adjust its volume and tone to match the room’s acoustics or blend with outdoor sounds. Users could switch between immersive sound experiences (where audio is more engaging) and ambient listening (where it remains subtle), ensuring technology subtly enhances the environment without drawing attention away from the primary activity.


Finally, AR could highlight real-world elements~\textcircled{\small{7}}—such as a sunset, the sound of crickets, or nearby landmarks—to enhance ongoing activities like listening to music or reading. By subtly drawing attention to these elements, AR can encourage users to engage with and appreciate their surroundings without disrupting their primary activity. For example, while reading outdoors, AR might amplify the gentle chirping of crickets to foster a calming atmosphere. Similarly, during a music session, AR could visually emphasise a vibrant sunset, aligning its hues with the music's tone to create a harmonious experience. To achieve this, AR systems must prioritise contextual awareness and personalisation. Understanding user activities, preferences, and environmental cues will enable AR to determine when to introduce enhancements. 

\begin{shaded}
\noindent AR can actively encourages reconnection with the physical world by integrating subtle cues that enhance mindfulness and awareness of the environment.
\end{shaded}

\section{Discussion}

\subsection{Designing AR for Presence and Engagement with Real World}

The findings of this study illustrate how AR may subtly guide attention, enhance presence, and mediate engagement with the real world. Rather than simply restricting digital use like existing digital well-being tools that focus on monitoring and limiting screen time~\cite{tiktok_screen_time, google_screen_time}, AR has the potential to proactively support meaningful interaction with the physical environment. 

Wearable AR, by leveraging context-awareness~\cite{grubert2017towards}, can offer subtle, personalised interventions to help users stay engaged with their surroundings. Inspired by nudging theory~\cite{thaler2009nudge}, AR can act as a digital ‘choice architect’, delivering timely, unobtrusive cues that gently encourage meaningful engagement. Unlike mobile notifications or screen-based nudges, AR can passively detect real-world context—such as user activity, gaze patterns, or environmental cues—to provide ambient guidance that blends seamlessly into the environment. As shown in the analysis, AR could introduce real-time prompts that encourage mindful breaks during passive scrolling or subtly adjust visual elements to facilitate natural transitions between digital and physical spaces. These interventions exemplify the principles of low-friction, non-intrusive design, integrating digital augmentation in ways that enhance rather than fragment real-world presence. Nevertheless, if not thoughtfully designed, AR could introduce risks that undermine real-world engagement rather than support it. Frequent prompts to take breaks or shift attention may become intrusive, while persistent overlays could contribute to cognitive overload. Over time, users may become desensitised to repetitive nudges, diminishing their effectiveness and turning them into background noise.

While AR can encourage reconnection with the real world, it is equally important to respect users’ autonomy and their freedom to choose intentional disengagement. Disengagement can serve as a valuable coping mechanism, allowing users to manage discomfort or take a break from social interactions. Drawing from nudging theory~\cite{thaler2009nudge} and a broad definition of nudges~\cite{bergram2022digitalnudges}, AR systems can embody `libertarian paternalism' by guiding users towards decisions that improve their well-being while preserving their freedom of choice. This approach avoids coercion, empowering individuals to navigate digital and physical interactions on their terms. Rather than framing disengagement as a negative behaviour to be discouraged, AR systems can balance escapism and situational awareness in ways that respect user autonomy. Subtle, user-driven interventions could maintain partial awareness of surroundings while allowing moments of retreat when desired. By enabling users to control their level of digital interaction, AR systems align with autonomy-supportive design principles, which prioritise user choice, self-determination, and individual needs~\cite{deci2012self}.

Balancing productivity and presence is a key challenge in designing AR systems. On one hand, AR can enhance productivity by automating content capture—such as detecting and saving key slides or generating real-time summaries—allowing users to stay focused on their surroundings instead of manually taking notes. However, over-reliance on automation risks reducing active engagement, where users passively consume information rather than processing it meaningfully. To address this, AR design must strike a balance, integrating automation in ways that support productivity without replacing active reflection with content.

\subsection{Autoethography as a Way of Generating Meaningful Use Cases for Everyday AR}

Autoethnography offers a unique lens for uncovering meaningful use cases for AR by grounding exploration in lived experiences and capturing the nuanced ways technology interacts with daily life. This reflective approach resonates with the informal methodology seen in \citet{glassner2003everyday}, where the author envisioned wearable AR assisting with small, casual tasks through intuitive, graphics-based, and text-free applications. While Glassner did not explicitly employ autoethnography, his use of personal experiences to derive practical scenarios highlights the value of reflective, experience-driven approaches in identifying everyday AR use cases. Similarly, this study demonstrates how documenting and reflecting on daily interactions can uncover subtle pain points, such as fragmented attention or the need for intentional disengagement, and suggest opportunities for AR to balance presence, productivity, and autonomy.

\subsection{Limitations and Future Work}

A key limitation of this study stems from its autoethnographic approach, which relies on personal, subjective reflections. While this method provides rich, nuanced insights into lived experiences, the findings may not generalise to broader populations or diverse user groups. Additionally, the analysis focuses primarily on interactions with existing technologies and may not fully capture the unique challenges posed by wearable AR devices. Their always-on and head-worn nature could introduce new forms of distraction and cognitive overload, which were not explored in this study. As such, the proposed design implications should be viewed as preliminary insights. Future research should adopt broader user-centred methods, including participatory design and iterative prototyping, to validate these directions and extend the findings to a wider range of contexts and user needs.

This study primarily focused on visual and auditory cues, similar approaches could extend to other sensory modalities, such as touch~\cite{wang2022ultrasonic} or scent~\cite{yanagida2019towards}. For example, haptic feedback could subtly reinforce attention shifts, while scent-based technologies could enhance spatial awareness or relaxation in certain environments. Future research could explore how multisensory AR interventions might further support presence and engagement by integrating these additional modalities.

Finally, this study defines presence simply as a person’s awareness of and engagement with their physical surroundings, in contrast to digital immersion. Meanwhile, in HCI literature, presence is often linked to virtual reality as the sense of `being there' a digital space~\cite{draper1998telepresence, witmer1998measuring}. Future research could explore individual notions of presence, as discussed by Terzimehić et al.~\cite{terzimehic2021ubiquitous, terzimehic2021real}, to better understand how people transition between digital and physical worlds.


\section{Conclusion}


This study explored how wearable AR can support real-world presence and engagement through context-aware and intentional design. Using an autoethnographic approach, it examined how everyday technologies shape presence and connection, identifying opportunities for wearable AR to balance digital and physical interactions. By challenging the notion of technology as inherently distracting, this research positions AR as a tool to enhance rather than disrupt everyday experiences. With thoughtful design, AR could not only reshape how we interact with technology but also how we reconnect with the world around us.


\bibliographystyle{ACM-Reference-Format}
\bibliography{references}

\appendix 

\section{Summary of Recorded Experiences}
\label{recorded-xp}

\begin{table*}[!ht]
\centering
\small
\caption{Phone Use During Meeting}
\label{tab:entry2_meeting}
\begin{tabular}{p{3cm}p{13cm}}
\toprule
\textbf{Date:} & \textbf{Wednesday, 13 November 2024} \\ 
\midrule
\textbf{Device/Application} & Phone \\ 
\textbf{Purpose} & Checking notifications, including a message from a friend regarding a pitching event in the evening. \\ 
\textbf{Context} & Attending a 2-hour meeting. \\ 
\textbf{Emotional Response} & Likely just a habit; I wanted to respond promptly to texts. \\ 
\textbf{Impact on Presence} & While taking meeting minutes, I missed parts of the discussion due to checking my phone. \\ 
\bottomrule
\end{tabular}
\end{table*}

\begin{table*}[!ht]
\centering
\small
\caption{Phone Use at Pitching Event}
\label{tab:entry3_pitching}
\begin{tabular}{p{3cm}p{13cm}}
\toprule
\textbf{Date:} & \textbf{Wednesday, 13 November 2024} \\ 
\midrule
\textbf{Device/Application} & Phone \\ 
\textbf{Purpose} & Browsing social media and emails. \\ 
\textbf{Context} & Attending a pitching event called Foundry; there is a networking session before the presentation. \\ 
\textbf{Emotional Response} & Socially tired and uninterested in talking to strangers. \\ 
\textbf{Impact on Presence} & Occasionally, I looked up to observe people talking to each other, engaging in human-watching. I also felt a bit self-conscious, hoping I didn’t appear awkward sitting alone, eating, and watching things. \\ 
\bottomrule
\end{tabular}
\end{table*}

\begin{table*}[!ht]
\centering
\small
\caption{Morning Phone Habit}
\label{tab:entry4_morning_phone}
\begin{tabular}{p{3cm}p{13cm}}
\toprule
\textbf{Date:} & \textbf{Sunday, 17 November 2024} \\ 
\midrule
\textbf{Device/Application} & Phone \\ 
\textbf{Purpose} & Browsing through different apps to check new emails, messages, news, and watch random videos. \\ 
\textbf{Context} & This felt like a habitual way to start the day. \\ 
\textbf{Emotional Response} & I felt overly involved with my phone, spending longer than needed. I wished I could avoid overwhelming myself with information the moment I wake up. \\ 
\textbf{Impact on Presence} & Aside from noticing the sunlight through the window and how my body felt, I didn’t pay attention to anything else while using my phone. I longed for a different morning routine—one with intentional activities like waking up, stretching, having a warm drink, and appreciating the room, the weather, and the season outside my window. \\ 
\bottomrule
\end{tabular}
\end{table*}

\begin{table*}[!ht]
\centering
\small
\caption{Phone Use During Breakfast}
\label{tab:entry5_breakfast}
\begin{tabular}{p{3cm}p{13cm}}
\toprule
\textbf{Date:} & \textbf{Sunday, 17 November 2024} \\ 
\midrule
\textbf{Device/Application} & Phone \\ 
\textbf{Purpose} & Reading a novel while eating breakfast (instant noodles and leftover meat from the previous night). \\ 
\textbf{Context} & Living alone, I enjoy eating while reading books or watching movies on my phone. \\ 
\textbf{Emotional Response} & I enjoy these moments a lot. \\ 
\textbf{Impact on Presence} & While some people prefer to focus solely on eating, I find joy in combining eating and reading. The two activities feel complementary rather than diminishing the value of each other. \\ 
\bottomrule
\end{tabular}
\end{table*}

\begin{table*}[!ht]
\centering
\small
\caption{Phone Use While Waiting for Coffee}
\label{tab:entry6_coffee_waiting}
\begin{tabular}{p{3cm}p{13cm}}
\toprule
\textbf{Date:} & \textbf{Sunday, 17 November 2024} \\ 
\midrule
\textbf{Device/Application} & Phone \\ 
\textbf{Purpose} & Continuing to read a novel while waiting for my coffee at the coffee store. \\ 
\textbf{Context} & Killing time and making the wait feel more interesting. \\ 
\textbf{Emotional Response} & Since the waiting time was short, I couldn’t immerse myself in the novel as I usually do. I also needed to pay attention to ensure my coffee was ready for pickup and to make way for customers entering the store. \\ 
\textbf{Impact on Presence} & I wasn’t fully present in either the novel I was reading or the reality around me. \\ 
\bottomrule
\end{tabular}
\end{table*}

\begin{table*}[!ht]
\centering
\small
\caption{Listening to Music While Walking}
\label{tab:entry7_music_walking}
\begin{tabular}{p{3cm}p{13cm}}
\toprule
\textbf{Date:} & \textbf{Sunday, 17 November 2024} \\ 
\midrule
\textbf{Device/Application} & Phone + Earphones \\ 
\textbf{Purpose} & Listening to music. \\ 
\textbf{Context} & I enjoy music and listen to it during my 15-minute walk to university. \\ 
\textbf{Emotional Response} & Listening to music has been a lifelong habit that I enjoy. \\ 
\textbf{Impact on Presence} & When I listen to music, I withdraw into a different world, almost like experiencing a `main character vibe,' as described in some funny social media videos. On campus, it feels safe due to low traffic. However, at intersections, I avoid changing the playlist to stay aware and avoid appearing distracted or delaying drivers. \\ 
\bottomrule
\end{tabular}
\end{table*}

\begin{table*}[!ht]
\centering
\small
\caption{Working with Music During Construction Noise}
\label{tab:entry9_work_noise}
\begin{tabular}{p{3cm}p{13cm}}
\toprule
\textbf{Date:} & \textbf{Monday, 18 November 2024} \\ 
\midrule
\textbf{Device/Application} & iMac and Earphones \\ 
\textbf{Purpose} & Getting work done while listening to music in the background. \\ 
\textbf{Context} & There is ongoing construction work next to my office at the university, making it difficult to focus without wearing earphones. The noise was very close and disruptive. \\ 
\textbf{Emotional Response} & I felt annoyed by the constant noise. Even when I used earphones, I had to turn up the volume, which I knew wasn’t good for my ears and felt too loud. \\ 
\textbf{Impact on Presence} & I wanted to withdraw completely from the outside world to escape the construction noise and focus on my work. \\ 
\bottomrule
\end{tabular}
\end{table*}

\begin{table*}[!ht]
\centering
\small
\caption{Watching Videos While Walking Home}
\label{tab:entry10_videos_walking}
\begin{tabular}{p{3cm}p{13cm}}
\toprule
\textbf{Date:} & \textbf{Monday, 18 November 2024} \\ 
\midrule
\textbf{Device/Application} & Phone + Earphones \\ 
\textbf{Purpose} & Watching short videos on Facebook while walking home. \\ 
\textbf{Context} & After finishing dinner and grocery shopping, I began my usual walk back home (a nearly 1 km walk). To make the walk more interesting, I started browsing short videos while wearing earphones. \\ 
\textbf{Emotional Response} & Initially, I was in a state of aimless scrolling, watching short videos without much thought. \\ 
\textbf{Impact on Presence} & As I neared home, I heard crickets chirping and looked up to see a glowing sunset in front of me. I took out my earphones and began to appreciate my surroundings—the sound of the crickets, the breeze, and the sunset. It made me reflect on what a nice day it had been. \\ 
\bottomrule
\end{tabular}
\end{table*}

\begin{table*}[!ht]
\centering
\small
\caption{Practising Japanese with ChatGPT While Walking}
\label{tab:entry8_japanese_chatgpt}
\begin{tabular}{p{3cm}p{13cm}}
\toprule
\textbf{Date:} & \textbf{Tuesday, 19 November 2024} \\ 
\midrule
\textbf{Device/Application} & Phone + Earphones \\ 
\textbf{Purpose} & Practising Japanese with ChatGPT through a speech-based conversation. \\ 
\textbf{Context} & I was taking a walk around my neighbourhood while practising Japanese with ChatGPT voice. During the conversation, it asked me: `Osanpo no aida, nani o mimasu ka?' (What do you see during your walk?). I replied, `Ame ga furisou desu.' (It looks like it’s going to rain). I then wanted to describe the grey sky but didn’t know the word in Japanese. \\ 
\textbf{Emotional Response} & I liked how natural the conversation felt, as if talking to a teacher or a friend. However, when ChatGPT provided the translation `Sora wa haiiro desu' (The sky is grey), I struggled to figure out the exact spelling of `haiiro'. Since it was a spoken interaction, I couldn’t visually confirm the word, which made me wish for a feature that could dynamically display the word in front of me, like an AR overlay. This would have helped me learn without interrupting the flow of the conversation. \\ 
\textbf{Impact on Presence} & While the speech-based interaction felt immersive, my focus shifted away from my physical surroundings as I tried to clarify the spelling. A tool that visually integrates words into the environment could enhance learning while maintaining awareness of the real-world context. \\ 
\bottomrule
\end{tabular}
\end{table*}

\begin{table*}[!ht]
\centering
\small
\caption{Using Phone to Wind Down Before Bed}
\label{tab:entry11_wind_down}
\begin{tabular}{p{3cm}p{13cm}}
\toprule
\textbf{Date:} & \textbf{20 November 2024} \\ 
\midrule
\textbf{Device/Application} & Phone \\ 
\textbf{Purpose} & Wind down before bed by browsing different things on my phone. \\ 
\textbf{Context} & I couldn’t sleep. \\ 
\textbf{Emotional Response} & I was physically tired but mentally still awake. I tried browsing things randomly to tire myself out and fall asleep. \\ 
\textbf{Impact on Presence} & I realised that just closing my eyes and lying still might have been more helpful than browsing aimlessly on my phone. The bedtime feature that turns the phone screen black and white at 9pm felt quite forced. The sudden switch from colour to black and white was jarring and did not feel natural. \\ 
\bottomrule
\end{tabular}
\end{table*}

\begin{table*}[!ht]
\centering
\small
\caption{Recording at a Concert}
\label{tab:entry12_concert_recording}
\begin{tabular}{p{3cm}p{13cm}}
\toprule
\textbf{Date:} & \textbf{26 November 2024} \\ 
\midrule
\textbf{Device/Application} & Phone \\ 
\textbf{Purpose} & Going to a concert and taking footage of some performances that I liked. \\ 
\textbf{Context} & Attended an outdoor concert. \\ 
\textbf{Emotional Response} & I had mixed feelings. I enjoyed recording short clips because they allowed me to relive the moment later. However, I also dislike the idea of people constantly using their phones during concerts, as it can feel like they are not truly living in the moment. \\ 
\textbf{Impact on Presence} & Recording did have some impact on my experience. It made it harder to dance or jump freely without worrying about shaking the phone. That said, I didn’t record every performance or for long periods, so it didn’t completely disrupt my presence. \\ 
\bottomrule
\end{tabular}
\end{table*}

\begin{table*}[!ht]
\centering
\small
\caption{Using Discord for Conference Communications}
\label{tab:entry13_discord_hci}
\begin{tabular}{p{3cm}p{13cm}}
\toprule
\textbf{Date:} & \textbf{1 December 2024} \\ 
\midrule
\textbf{Device/Application} & Phone \\ 
\textbf{Purpose} & Checking new messages on Discord channels. \\ 
\textbf{Context} & I used Discord to communicate with the Student Volunteers team during an HCI conference. \\ 
\textbf{Emotional Response} & I usually keep my phone on silent mode to avoid distractions, but during social and management tasks like this, it was hard to balance staying undistracted while ensuring I didn’t miss important communications. \\ 
\textbf{Impact on Presence} & While my phone remained on silent mode, I found myself glancing at it more often throughout the day, which affected my ability to remain fully present. \\ 
\bottomrule
\end{tabular}
\end{table*}

\begin{table*}[!ht]
\centering
\small
\caption{Fitbit Low Battery and Tracking Disruption}
\label{tab:entry14_fitbit_steps}
\begin{tabular}{p{3cm}p{13cm}}
\toprule
\textbf{Date:} & \textbf{1 December 2024} \\ 
\midrule
\textbf{Device/Application} & Fitbit \\ 
\textbf{Purpose} & Tracking daily steps. \\ 
\textbf{Context} & The device ran out of battery, but I was not aware. This led to a loss of several days’ worth of step-tracking data. \\ 
\textbf{Emotional Response} & I felt quite annoyed about losing data. \\ 
\textbf{Impact on Presence} & It’s nice that the Fitbit functions unobtrusively without demanding attention, but it would be helpful if it could notify me when important issues, such as low battery, occur. This could help balance its quiet operation with timely, necessary alerts. \\ 
\bottomrule
\end{tabular}
\end{table*}

\begin{table*}[!ht]
\centering
\small
\caption{Using Phone During a Presentation}
\label{tab:entry15_conference_presentation}
\begin{tabular}{p{3cm}p{13cm}}
\toprule
\textbf{Date:} & \textbf{4 December 2024} \\ 
\midrule
\textbf{Device/Application} & Phone \\ 
\textbf{Purpose} & Googling information and taking photos of important slides. \\ 
\textbf{Context} & While attending a presentation at a conference. \\ 
\textbf{Emotional Response} & It felt like a necessary thing to do at the time. However, I also recognised that I rarely look back at these photos later because they get buried in the gallery, making it hard to revisit them. \\ 
\textbf{Impact on Presence} & Using the phone was slightly distracting. It took time to Google information and capture slides, which pulled my attention away from the presentation. Occasionally, I wasn’t fast enough, and the speaker had already moved on to the next slide before I could take a photo. \\ 
\bottomrule
\end{tabular}
\end{table*}

\begin{table*}[!ht]
\centering
\small
\caption{Multi-Tasking During an Event}
\label{tab:entry16_event_notes}
\begin{tabular}{p{3cm}p{13cm}}
\toprule
\textbf{Date:} & \textbf{6 December 2024} \\ 
\midrule
\textbf{Device/Application} & Phone \\ 
\textbf{Purpose} & Writing notes on the phone during an event, searching for people’s names, and connecting with them on LinkedIn. \\ 
\textbf{Context} & At the VisionOS Pro event. \\ 
\textbf{Emotional Response} & The experience involved multi-tasking—I was taking notes on interesting points during the event, but I lost track of what the speakers were saying next. This created a sense of disconnection and fragmented focus. \\ 
\textbf{Impact on Presence} & My engagement with the event was very short and temporary. While the phone was useful for note-taking and networking, it distracted me from the ongoing presentation and made it harder to stay present in the moment. \\ 
\bottomrule
\end{tabular}
\end{table*}

\begin{table*}[!ht]
\centering
\small
\caption{ Watching Chess Commentary Across Devices}
\label{tab:entry17_chess_commentary}
\begin{tabular}{p{3cm}p{13cm}}
\toprule
\textbf{Date:} & \textbf{12 December 2024} \\ 
\midrule
\textbf{Device/Application} & Phone + Laptop \\ 
\textbf{Purpose} & Watching live-streamed commentary of the FIDE chess championship match. \\ 
\textbf{Context} & I used multiple tabs on my laptop to watch different live commentary streams. The FIDE official stream featured an evaluation bar showing win/draw percentages but I did not enjoy the commentary. ChessBase India provided exciting commentary, often featuring guest speakers joining via calls or Zoom, but its simplified evaluation bar felt less useful. I also used the Take Take Take app on my phone for short written updates, visualised engine moves, and more structured analysis. \\ 
\textbf{Emotional Response} & I felt slightly overwhelmed by the scattered information across tabs and devices.  \\ 
\textbf{Impact on Presence} & Switching between these platforms fragmented my experience. Managing multiple streams and devices distracted me from focusing fully on the game. While the tools offered different perspectives, they collectively took away the opportunity to appreciate the game’s beauty. \\ 
\bottomrule
\end{tabular}
\end{table*}

\begin{table*}[!ht]
\centering
\small
\caption{Classical Music Concert Experience}
\label{tab:entry18_classical_concert}
\begin{tabular}{p{3cm}p{13cm}}
\toprule
\textbf{Date:} & \textbf{14 December 2024} \\ 
\midrule
\textbf{Device/Application} & Phone \\ 
\textbf{Purpose} & Taking a photo of the pianist at the end of the concert and capturing the beauty of the concert hall. \\ 
\textbf{Context} & The event was a classical music concert where the audience was not supposed to use phones to maintain silence and avoid distracting the pianist or others. \\ 
\textbf{Emotional Response} & I really liked how we were able to focus and appreciate every single note, especially the very low and subtle ones. Hearing how the sounds travelled within the concert hall felt magical and immersive. \\ 
\textbf{Impact on Presence} & I felt totally immersed in the performance. However, the lack of lyrics and the absence of listed encores in the program made me mentally distracted during the first encore. I found myself trying to memorise the melody or guess the composer and piece type (e.g., nocturne, ballade, etude, polonaise) so I could search for it later at home. Despite not using my phone, I was still slightly distracted by this mental effort. \\ 
\bottomrule
\end{tabular}
\end{table*}

\begin{table*}[!ht]
\centering
\small
\caption{Listening to Music While Reading}
\label{tab:entry19_music_reading}
\begin{tabular}{p{3cm}p{13cm}}
\toprule
\textbf{Date:} & \textbf{15 December 2024} \\ 
\midrule
\textbf{Device/Application} & Phone \\ 
\textbf{Purpose} & Listening to music. \\ 
\textbf{Context} & I was reading books and wanted some background music. Normally, I would use earphones, but this time I used the phone’s speaker to play Southern All Stars' "Manatsu no Kajitsu". \\ 
\textbf{Emotional Response} & It was hard to be totally immersed in the music without earphones since the sound quality felt reduced. However, the experience gave me a nostalgic feeling, similar to listening to the radio during my childhood. This brought a subtle sense of comfort. \\ 
\textbf{Impact on Presence} & Presence in what? I didn’t feel fully present in the music itself, but I was present in the real world, with music playing softly in the background. The music created a light, ambient atmosphere that accompanied my reading without completely pulling me in. \\ 
\bottomrule
\end{tabular}
\end{table*}

\begin{table*}[!ht]
\centering
\small
\caption{Reading and Music at the Laundromat}
\label{tab:entry20_reading_breeze}
\begin{tabular}{p{3cm}p{13cm}}
\toprule
\textbf{Date:} & \textbf{16 December 2024} \\ 
\midrule
\textbf{Device/Application} & Laptop + Earphones \\ 
\textbf{Purpose} & Reading a research paper while listening to music. \\ 
\textbf{Context} & I was reading a research paper on my laptop while waiting for the laundry to finish at a neighbourhood laundromat. It’s summer, and the front door was open. While reading, I felt a cool breeze, looked out at the traffic on the road, and appreciated how refreshing the summer breeze felt. \\ 
\textbf{Emotional Response} & I felt pretty happy. The combination of music, reading, and the small moment of noticing the breeze brought unexpected joy. \\ 
\textbf{Impact on Presence} & The cool breeze momentarily took me away from reading the paper. While it interrupted my focus, it was a welcome, pleasant break that allowed me to notice and appreciate my surroundings. \\ 
\bottomrule
\end{tabular}
\end{table*}

\end{document}